# Resolving a Candidate Dual Active Galactic Nucleus with ∼100 pc Separation in MCG-03-34-64

Anna Trindade Falcão[1] , T. J. Turner[2] , S. B. Kraemer[3] , J. Reeves[3,4] , V. Braito[3,4,5] , H. R. Schmitt[6] , and L. Feuillet[3]
[1] Harvard-Smithsonian Center for Astrophysics, 60 Garden St., Cambridge, MA 02138, USA
[2] Eureka Scientific, Inc., 2452 Delmer St., Suite 100, Oakland, CA 94602, USA
[3] Institute for Astrophysics and Computational Sciences, The Catholic University of America, Washington, DC 20064, USA
[4] INAF—Osservatorio Astronomico di Brera, Via Bianchi 46, 23807, Merate (LC), Italy
[5] Dipartimento di Fisica, Università di Trento, Via Sommarive 14, Trento 38123, Italy
[6] Naval Research Laboratory, Washington, DC 20375, USA

## Abstract

We report the serendipitous multiwavelength discovery of a candidate dual black hole system with a separation of ∼100 pc, in the gas-rich luminous infrared galaxy MCG-03-34-64 ($z = 0.016$). Hubble Space Telescope/Advanced Camera for Surveys observations show three distinct optical centroids in the [O III] narrow-band and F814W images. Subsequent analysis of Chandra/ACIS data shows two spatially resolved peaks of equal intensity in the neutral Fe Kα (6.2–6.6 keV) band, while high-resolution radio continuum observations with the Very Large Array at 8.46 GHz (3.6 cm band) show two spatially coincident radio peaks. Fast shocks as the ionizing source seem unlikely, given the energies required for the production of Fe Kα. If confirmed, the separation of ∼100 pc would represent the closest dual active galactic nuclei reported to date with spatially resolved, multiwavelength observations.

*Unified Astronomy Thesaurus concepts:* AGN host galaxies (2017); Seyfert galaxies (1447); High energy astrophysics (739)

## 1. Introduction

The masses of supermassive black holes (SMBHs) in active galactic nuclei (AGNs) correlate with the global properties of their host galaxies' stellar components, such as luminosity, mass, and velocity dispersion, extending over kiloparsec scales (e.g., Kormendy & Ho 2013). This correlation highlights the need to understand the mechanisms driving SMBH growth.

Both galactic evolutionary models and observations suggest that a significant fraction of AGNs, particularly those at the center of large-scale structures, undergo major mergers (e.g., De Lucia & Blaizot 2007; Hopkins et al. 2008; Ginolfi et al. 2017; Castignani et al. 2020). Hydrodynamical simulations further demonstrate that major mergers induce gas inflows toward galactic centers, potentially triggering both star formation and accretion onto central SMBHs (Mayer et al. 2007). However, the overall impact of these events on SMBH growth throughout cosmic time remains poorly constrained.

SMBH pairs, often manifested as dual AGNs, provide distinctive evidence for merger-fueled SMBH growth (e.g., Wassenhove et al. 2012). Numerous dual AGN candidates have been identified using various techniques, including optical spectroscopy with emission line ratios (e.g., Liu et al. 2011), hard X-ray emission (e.g., Koss et al. 2011), and double-peaked narrow emission lines (e.g., Smith et al. 2010; Koss et al. 2023). Nonetheless, these methods have limitations, and multiwavelength follow-up observations have revealed a substantial number of false positives (e.g., Fu et al. 2011b).

The advent of gravitational-wave astronomy, with the potential for detection through pulsar timing arrays (e.g., Verbiest et al. 2016), has heightened the importance of understanding the formation timescales of binary systems. Studying kiloparsec and subkiloparsec dual AGNs offers a unique window into the final stages of SMBH binary coalescence, a crucial process in gravitational wave astronomy.

Dual AGNs separated by kiloparsec or subkiloparsec scales are inherently more challenging to detect and investigate than wider-separation systems (e.g., >3 kpc). This difficulty arises from increased obscuration in late-stage mergers (e.g., Koss et al. 2016; Ricci et al. 2021; De Rosa et al. 2022), limitations in telescope spatial resolution (particularly at subkiloparsec scales), the scarcity of detected radio-bright dual systems (Burke-Spolaor 2011), and the limitations of optical selection using double-peaked narrow emission lines (prone to false positives; see Fu et al. 2011a). Existing observations of dual AGNs tentatively suggest that AGN triggering becomes more prevalent in advanced mergers with stellar bulge separations <10 kpc (e.g., Koss et al. 2010; Fu et al. 2018; Stemo et al. 2021), aligning with simulations of SMBH accretion and evolution in such mergers (e.g., Blecha et al. 2018). Therefore, studying nearby galaxies hosting dual AGNs separated at subkiloparsec scales is crucial for advancing our understanding of the late stages of galaxy mergers, the triggering and fueling of AGN activity, and the dynamics of SMBH pairs (Steinborn et al. 2016). These close-separation systems provide a unique window into the processes leading to the eventual coalescence of SMBHs, which is a major source of gravitational waves, and plays a fundamental role in the growth of SMBHs and their host galaxies (Dotti et al. 2012; Kharb et al. 2017).

While several dual AGN candidates have been proposed at scales of hundreds of parsecs, often supported by single-waveband observations, these have frequently been challenged by subsequent studies. Notable examples include the nearby Seyfert NGC 3393 (Fabbiano et al. 2011), SDSS J101022.95 + 141300.9

Table 1
Multiwavelength Observations of MCG-03-34-64

| Wavelength Band | Instrument/Filter | Date of Observation | Observation ID | Exposure Time (s) |
|---|---|---|---|---|
| Optical | [a]HST/ACS FR505N | 2022-06-30 | jequ01020 | $1.5 \times 10^3$ |
| | [a]HST/ACS FR647M | 2022-06-30 | jequ01010 | $2.0 \times 10^2$ |
| | HST/ACS F814W | 2019-01-18 | jdrw9z010 | $7.0 \times 10^2$ |
| Radio | VLA/A (8.46 GHz) | 1995-07-15 | AK394 | $9.0 \times 10^2$ |
| X-rays | [a]Chandra/ACIS-S | 2023-04-19 | 25253 | $1.5 \times 10^4$ |
| | [a]Chandra/ACIS-S | 2023-04-20 | 27802 | $1.8 \times 10^4$ |
| | [a]Chandra/ACIS-S | 2023-04-21 | 27803 | $1.7 \times 10^4$ |
| | NuSTAR | 2009-07-01 | 60101020002 | $7.8 \times 10^4$ |
| | XMM-Newton/Epic-pn | 2016-01-17 | 0763220201 | $1.0 \times 10^5$ |

**Note.**
[a] New observations.

(Goulding et al. 2019), and a third active nucleus in NGC 6240 (Kollatschny et al. 2020), later disputed in other works (Koss et al. 2015; Veres et al. 2021; Treister et al. 2020).

Recently, Koss et al. (2023) have used multiwavelength observations to confirm the presence of two active nuclei separated by ∼230 pc in the center of UGC 4211, a late-stage major galaxy merger. Their discovery builds on the detection of (i) near-infrared (NIR) broad lines associated with one of the nuclei; (ii) two separated sources of [O III] emission, unresolved at ∼0″.1 resolution; (iii) unresolved and compact (<30 pc) nuclear millimeter emission at the location of the two nuclei, and coincident with the unresolved emission-line sources; and (iv) molecular gas velocity gradient centered on the position of both nuclei.

In this study, we present the serendipitous discovery of a candidate dual AGN system in MCG-03-34-64 (IRAS 13197-1627), a nearby early-type infrared luminous galaxy at $z = 0.01654$ (∼78 Mpc, from NASA/IPAC Extragalactic Database).[7] This galaxy is identified as one of the hardest X-ray sources in the local Universe (Tatum et al. 2016). Earlier X-ray observations with ASCA, XMM-Newton, and BeppoSAX (Dadina & Cappi 2004; Miniutti et al. 2007) revealed an extremely hard and complex source spectrum, attributed to heavy absorption from a multilayered and clumpy medium. MCG-03-34-64 also shows extended radio emission (∼300 pc), roughly aligned with the major axis of the host galaxy (Schmitt et al. 2001), and ∼2″ extent in mid-infrared aligned in the same direction as the radio structure (Hönig et al. 2010).

We have obtained Hubble Space Telescope/Advanced Camera for Surveys (ACS) imaging of MCG-03-34-64 in 2022 June (P.I.: Turner, proposal ID: 16847), and 50 ks of Chandra/ACIS-S observations in 2023 April (obs ids 25253, 27802, and 27803, P.I.: Turner). This paper presents the results of the analysis of these new data sets, combined with existing Very Large Array (VLA) radio, and Hubble Space Telescope (HST) optical imaging of the source. Throughout this paper, we adopt $\Omega_m = 0.3$, $\Omega_\Lambda = 0.7$, and $H_0 = 70 \text{ km s}^{-1} \text{ Mpc}^{-1}$, and a scale of 340 pc arcsec$^{-1}$, based on the redshift at the galaxy's distance.

## 2. Multiwavelength Observations and Analysis

Table 1 lists all observations used in this paper, including instruments, filters, observation dates, obs ids, and exposure times. Details on the reduction of new HST/ACS and Chandra/ACIS-S observations are provided in Sections 2.1 and 2.2, respectively. For reduction and analysis of archival HST F814W, VLA 8.46 GHz imaging, and Suzaku and XMM-Newton spectroscopy, see Section 2.3. All the HST data used in this paper can be found in MAST at doi:10.17909/53rj-fw34.

### 2.1. Hubble Space Telescope Imaging

HST/ACS observations of MCG-03-34-64 were obtained using the linear ramp filter FR505N (narrow-band [O III]) centered at 5089.6 Å, to characterize the morphology of the emission-line gas, while a continuum medium band image was obtained using FR647M, centered at 5590 Å. These filters have bandwidths of 2% and 9%, respectively. Standard HST pipeline procedures were employed for data reduction. The narrow-band and continuum images were acquired sequentially and did not require realignment. Flux calibration was performed using information available on the headers.

### 2.2. Chandra Imaging and Spectroscopy.

MCG-03-34-64 was observed with Chandra/ACIS for a total exposure of 50 ks, split into three observations (Table 1). We use CIAO 4.15[8] for data processing and analysis, merging individual observations to create a deeper data set for imaging analysis. Each image in the 0.3–7 keV band was first visually inspected, and the required centroid shifts were found to be less than one ACIS instrument pixel (0″.492). We created images in the 2–7 keV band, and using obs id 27802 (longest observation, see Table 1) as the astrometric reference, shifts of [$\Delta x$, $\Delta y$] = [−0.7,−0.8], and [0.8,0.3] pixels were applied to obs ids 25253 and 27803, respectively. All data sets were processed to enable subpixel analysis (e.g., Wang et al. 2011a).

Subpixel imaging binning was employed to effectively oversample the Chandra point-spread function (PSF) and overcome the limitations of the ACIS instrumental pixel size. This method has been extensively used and validated in previous studies examining the subkiloparsec regions around nearby and obscured AGNs (e.g., Maksym et al. 2017; Fabbiano et al. 2018a; Ma et al. 2021; Trindade Falcão 2023), demonstrating excellent agreement between reconstructed ACIS-S features and those imaged with higher spatial resolution instruments such as HST and VLA (e.g., Wang et al. 2011b; Paggi et al. 2012; Maksym et al. 2019; Fabbiano et al. 2018b). The Chandra PSF was simulated using ChaRT[9] and MARX.[10] This work uses a final Chandra scale of one-eighth of the native ACIS pixel.

---

[7] https://ned.ipac.caltech.edu/

[8] https://cxc.cfa.harvard.edu/ciao/

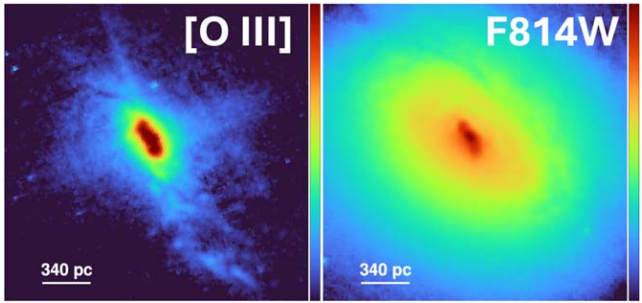

**Figure 1.** HST [O III] and F814W images of the central region of MCG-03-34-64. Note the prominent diffraction spikes present in the [O III] image.

For the spectral analysis (Section 3.2), we used CIAO specextract[11] to extract the spectra from a $2''\!.5$ circular region, while the background was subtracted from a circular region of $3''\!.5$ radius. Individual spectra and response files were coadded with mathpha[12] and the standard addrmf[13] and addarf[14] functions.

### 2.3. Archival Radio/Optical/X-Ray Observations

In addition to the new Chandra and HST data sets, we analyze archival optical, radio, and X-ray observations of MCG-03-34-64, as listed in Table 1. These data include 8.46 GHz radio imaging with VLA, optical continuum imaging with HST/ACS F814W, and X-ray spectra from Suzaku and XMM-Newton.

There are four additional archival Chandra observations with MCG-03-34-64 in the field of view (obs ids 27267, 27786, 7373, and 23690). However, three of these observations are not usable due to the galaxy being located at the very edge of the field (observations were optimized for the companion galaxy). The fourth available Chandra observation consists of a 7 ks snapshot (used in Miniutti et al. 2007), which has insufficient counts for meaningful imaging analysis.

## 3. Results

### 3.1. Imaging Analysis

#### 3.1.1. Hubble Space Telescope Imaging

The [O III] narrow-line region (NLR) in MCG-03-34-64 has a highly unusual morphology, featuring three distinct, and compact emission regions, as shown in Figure 1. The NLR extends ∼2.3 kpc along the NE–SW direction. In the perpendicular direction (NW–SE), we observe three diffraction spikes characteristic of point sources, while one diffraction spike is observed along the NE–SW cone. These features suggest high concentrations of [O III] gas within a relatively small region, a rare occurrence in the local Universe (Fischer et al. 2018).

---
[9] https://cxc.cfa.harvard.edu/ciao/PSFs/chart2/
[10] https://space.mit.edu/cxc/marx/
[11] https://cxc.cfa.harvard.edu/ciao/ahelp/specextract.html
[12] https://heasarc.gsfc.nasa.gov/lheasoft/ftools/fhelp/mathpha.html
[13] https://heasarc.gsfc.nasa.gov/lheasoft/ftools/fhelp/addrmf.html
[14] https://heasarc.gsfc.nasa.gov/lheasoft/ftools/fhelp/addarf.html

To examine the overall structure of the NLR, we model the [O III] light distribution with GALFIT (Peng et al. 2002, 2010). The modeling was performed with generic Sérsic profiles, and we allow all parameters to freely vary during the fitting process. The background level and standard deviation were determined from blank regions within the image's field of view.

The best-fit GALFIT model consists of four Sérsic components, one for each of the three peaks visible in the [O III] surface plot in the leftmost panel of Figure 2, and an additional larger component to take into account the underlying fainter, more extended emission. In the second panel, we show a zoomed-in image of the [O III] emission, and mark the position of each Sérsic component corresponding to a strong peak of [O III] emission as green circles. The third panel of Figure 2 shows the best-fit GALFIT model, while the fourth panel shows the residual images from this best-fit model, with the different model components subtracted from the data.

Table 2 lists the best-fit model parameters. The strong [O III] peaks of emission are fit with Sérsic components with indices $0.41 \leqslant n \leqslant 0.51$, indicating that they are similar to Gaussians ($n = 0.5$), but in two cases show a slightly more centrally concentrated distribution. Given that these components have effective radii ∼2–3 times that of an unresolved source, we opted to not include a PSF component in the fitting model, since it is not expected to significantly change the results.

The diffraction spikes associated with these regions (Figure 1) suggest angular sizes on the order of HST/ACS's resolution element, $0''\!.1$ (2 pixels). Assuming the measured effective radius, $R_e$, of each [O III] component in Table 2 represents its intrinsic size, we estimate that each [O III]-emitting region has an average angular diameter of ∼$0''\!.2$, or 60 pc. These regions are separated by $79 \pm 8$ pc (northern to central) and $76 \pm 8$ pc (central to southern), and are also clearly observed in the broad continuum F814W HST image (Figure 1, right panel), a point further discussed in Section 4.

Given the unique morphology of the optical emission observed in MCG-03-34-64 with HST, we examine other available multiwavelength observations (Table 1) to obtain a more comprehensive picture of this inner region.

#### 3.1.2. Chandra Imaging

Figure 3 shows the inner ∼200 pc region of MCG-03-34-64, as observed with Chandra/ACIS in different energy bands. The images are binned at one-eighth of the native pixel to use the full-resolution of the instrument in the high-count inner region, and processed with 1 kernel Gaussian smoothing.

The X-ray-emitting region shows a compact morphology, with the most prominent features observed within the inner ∼$0''\!.5$ (∼170 pc) circumnuclear region. Extended diffuse emission is detected in all bands, as shown in Figure 3; however, there are clear energy-dependent differences in the surface brightness morphology. At higher energies (Figure 3, two rightmost panels), the emission from the nuclear region cannot be explained as a single point source. Instead, in the hard neutral Fe K$\alpha$ (6.2–6.6 keV) and 6–7 keV bands, we observe two spatially resolved peaks of emission separated by $125 \pm 21$ pc.

We measure nearly equal X-ray luminosities in the narrow 6.2–6.6 keV band from the Chandra image for both Fe K$\alpha$ peaks, $L_{(6.2-6.6 \text{ keV})} \sim 3.2 \pm 0.6 \times 10^{40}$ erg s$^{-1}$, for $D \sim 78$ Mpc (e.g., de Grijp et al. 1992; Table 3).

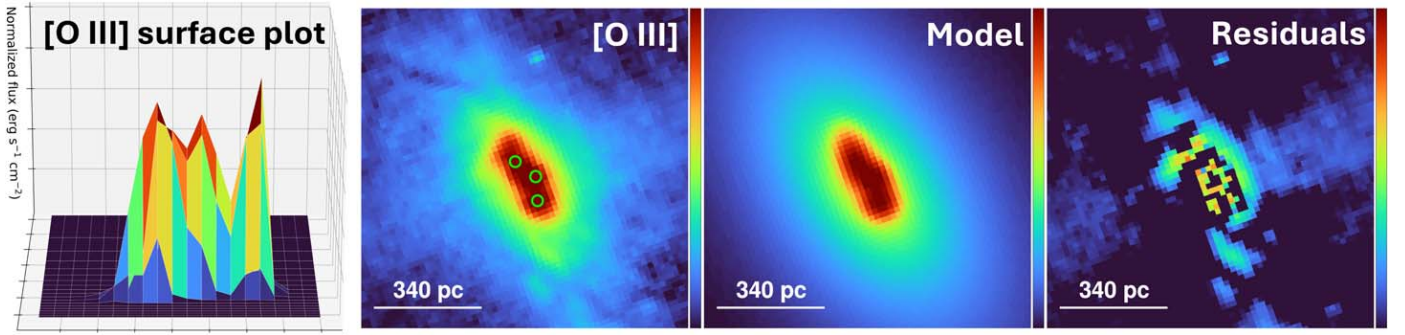

**Figure 2.** First panel: surface plot of the [O III] flux in the HST/ACS image where three distinct sources are visible. Second panel: HST/ACS [O III] image showing the presence of three closely separated emission peaks. We mark the position of each Sérsic component for the emission peaks as green circles. Third panel: best-fit `GALFIT` model, using four Sérsic components. Fourth panel: residual image after subtraction of the best-fit `GALFIT` model.

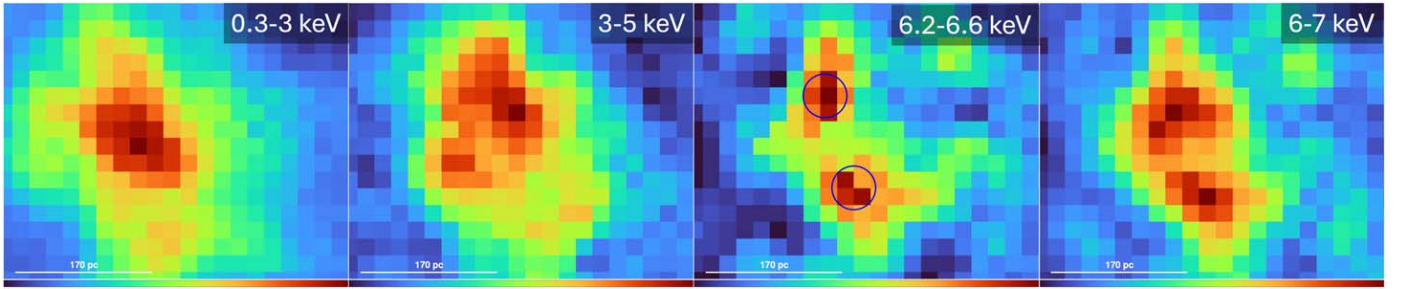

**Figure 3.** Chandra/ACIS-S merged data set showing the inner 200 pc region of MCG-03-34-64 in different bands (one-eighth subpixel, and 1 kernel Gaussian smoothing). First panel: soft (0.3–3 keV) X-ray image. Second panel: 3–5 keV hard continuum image. Third panel: Fe K$\alpha$ (6.2–6.6 keV) Chandra image. We show the location of the two Fe K$\alpha$ centroids as blue circles. Fourth panel: 6–7 keV hard band Chandra image.

**Table 2**
Best-fit `GALFIT` Parameters ($\chi^2_\nu = 5.411$)

| Component | R.A. (J1200) | Decl. (J1200) | $R_e$ (arcsec) | $n$ | $b/a$ | P.A. (sky) (deg) | Flux (erg s$^{-1}$ cm$^{-2}$) |
|---|---|---|---|---|---|---|---|
| Sérsic I (northern [O III]) | 13:22:24.4693 | −16:43:42.379 | 0.095 | 0.41 | 0.92 | 49.9 | $4.8 \times 10^{-13}$ |
| Sérsic II (central [O III]) | 13:22:24.4561 | −16:43:42.515 | 0.101 | 0.51 | 0.64 | 32.3 | $2.2 \times 10^{-13}$ |
| Sérsic III (southern [O III]) | 13:22:24.4549 | −16:43:42.737 | 0.160 | 0.44 | 0.53 | 24.9 | $3.2 \times 10^{-13}$ |
| Sérsic IV (fainter extended [O III]) | 13:22:24.4625 | −16:43:42.576 | 0.508 | 2.91 | 0.59 | 34.3 | $4.6 \times 10^{-13}$ |

**Table 3**
Chandra/ACIS-S Merged Data Set—Astrometry, Fluxes, and Luminosities of the Individual Fe K$\alpha$ Regions

| Component | R.A. (J1200) | Decl. (J1200) | Counts (photons) | Flux (erg s$^{-1}$ cm$^{-2}$) | Luminosity (erg s$^{-1}$) |
|---|---|---|---|---|---|
| Northern Fe K$\alpha$ | 13:22:24.470 | −16:43:42.314 | $36 \pm 6$ | $4.6 \times 10^{-14}$ | $3.2 \pm 0.6 \times 10^{40}$ |
| Central Fe K$\alpha$ | 13:22:24.4584 | −16:43:42.643 | $37 \pm 6$ | $4.8 \times 10^{-14}$ | $3.2 \pm 0.6 \times 10^{40}$ |

**Note.** The luminosities are calculated for $D = 78$ Mpc.

To address concerns that the dual morphology might be a spurious detection due to smoothing on scales smaller than the Chandra PSF, we examine the individual Chandra observations (prior to merging) listed in Table 1, in the Fe K$\alpha$ band (Figure 4). These images are binned at one-eighth of the native pixel and smoothed with a 1 kernel Gaussian. As shown, the dual morphology of the Fe K$\alpha$ band is indeed observed in all individual observations, prior to merging, confirming the robustness of the detection. The differences in surface brightness between the individual Chandra exposures seen in Figure 4 for the individual nuclei are most likely due to statistical noise.

### 3.1.3. VLA Imaging

The morphology of the 8.46 GHz radio continuum emission in MCG-03-34-64 is also analyzed, and shown in Figure 5. The positions of the two radio centroids identified by Schmitt et al. (2001) in the inner region are shown as white circles (see their Table 2). The 8.46 GHz emission starts as a linear structure at the position of the northern radio centroid, extending ∼100 pc southwestward to the central radio peak, and then bending southward in the direction of the southern [O III] centroid (Figure 5). Table 4 lists the positions and fluxes of the individual

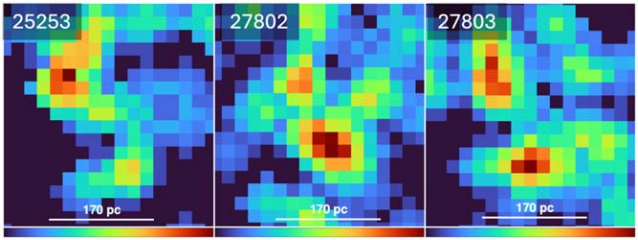

**Figure 4.** Chandra/ACIS individual exposures of MCG-03-34-64 in the Fe Kα (6.2–6.6 keV) band, binned at one-eighth of the native pixel and smoothed with a 1 kernel Gaussian. As shown, the dual morphology is observed prior to merging. The distances between the two peaks in the individual observations are 0″486, 0″408, and 0″443, for obs ids 25253, 27802, and 27803, respectively. Given the 0″08 uncertainty in the Chandra data, these are all consistent within the error.

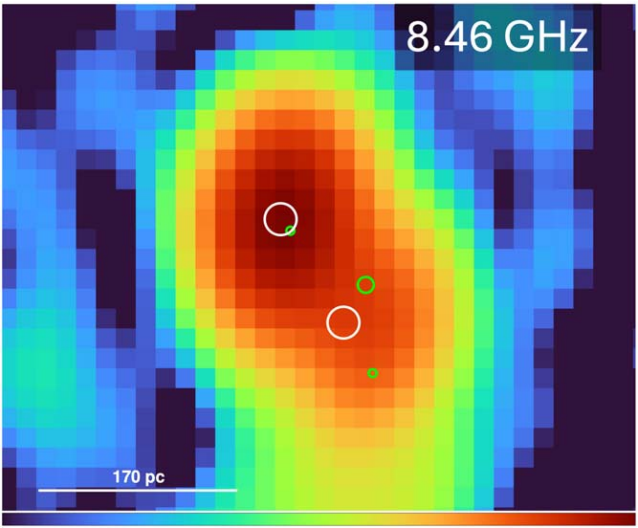

**Figure 5.** VLA-A 8.46 GHz (3.6 cm) radio continuum image of MCG-03-34-64. White circles mark the position of the two radio peaks, and the [O III] centroids are shown in green. The image is shown in log scale.

**Table 4**
VLA-A 3.6 cm Radio Continuum Image Decomposition—Position and Fluxes of Individual Components

| Component | R.A. (J1200) | Decl. (J1200) | Flux (mJy) |
|---|---|---|---|
| Northern radio | 13:22:24.471 | −16:43:42.35 | 31.6 |
| Central radio | 13:22:24.456 | −16:43:42.61 | 15.5 |

**Note.** From Schmitt et al. (2001).

radio components, as measured in Schmitt et al. (2001). The separation between the two centroids is $116 \pm 14$ pc.

### 3.1.4. Astrometry Registration

We initially apply CIAO wavdetect[15] to the merged 0.3–7 keV Chandra image with a >5σ detection threshold, detecting two faint sources near the edge of the chip array (via comparison with the Vizier source catalog). However, given their faintness, these are not suitable to use as a basis for astrometry correction.

We then create Chandra images in the narrow Fe Kα band (6.2–6.6 keV), known to be dominated by nuclear emission in obscured sources (Fabbiano & Elvis 2024). Assuming that the radio emission in AGNs also originates from the innermost regions around the SMBH, we correct Chandra's absolute astrometry by aligning the emission peaks seen in the Fe Kα band with those in the 8.46 GHz radio image. This alignment method has been used in similar studies, such as in Mrk 78 (Fornasini et al. 2022).

We apply a total shift of $[\Delta x, \Delta y] = [0.1, 0.7]$ pixels to the Chandra/ACIS data, well within the absolute astrometry accuracy of the telescope.[16] Comparison with available HST observations confirms the accuracy of VLA's astrometry.

### 3.2. Spectroscopic Analysis

We also analyzed Chandra/ACIS, NuSTAR, and XMM-Newton spectroscopic observations of MCG-03-34-64, spanning 14 yr (Table 1). In Figure 6 (left panel), we show the combined 50 ks Chandra/ACIS spectrum of the source, which was extracted from a 2″5 circular region. This spectrum is heavily absorbed at the lower energies, below 5 keV, indicating a heavily obscured nuclear source. At higher energies, we observe a broad neutral Fe Kα line, with $E = 6.39 \pm 0.01$ keV (the 6.2–6.6 keV Fe Kα band is shown as red lines), equivalent width $EW = 470 \pm 60$ eV, and $\sigma = 0.11 \pm 0.05$ keV.

The overall spectral profile observed in the Chandra data is consistent across the earlier NuSTAR and XMM-Newton observations. All three spectra show an absorption feature at 6.8 keV, likely arising from Fe XXV absorption, and suggesting an outflow with $v \sim 5000$ km s$^{-1}$, as noted in Miniutti et al. (2007). Below 3 keV, the soft X-ray emission is dominated by photoionized and collisionally ionized emission lines, from Ne, O, and Fe L ions (Miniutti et al. 2007).

#### 3.2.1. XMM-Newton and NuSTAR Spectral Fitting

We proceed to model the X-ray spectrum of MCG-03-34-64 with MYTorus (Murphy & Yaqoob 2009), a physically motivated model built to describe the interaction of the emission from an X-ray point-source with a surrounding, and homogeneous torus of cold neutral material.

We fit the joint XMM+NuSTAR X-ray spectrum with a source model of the form:

A × TBabs × [xstar × MYTZ × zpowerlw+(C × (MYTS+ MYTL × gsmooth)+zpowerlaw_soft+soft_emiss)],

where TBabs describes the absorption of emission by the Galactic column density, xstar is the photoionized absorber described previously in Miniutti et al. (2007), [MYTZ × zpowerlw] describes the intrinsic continuum in transmission absorbed by torus, MYTS is the scattered (reflected) toroidal component off Compton-thick matter, MYTL is the associated Fe/Ni Kα/Kβ line emission, and gsmooth accounts for some Gaussian broadening of the MYTorus line emission, where the upper limit on the line width is $\sigma < 65$ eV. The soft X-ray components are zpowerlaw_soft, which is an unabsorbed scattered power-law component, and soft_emiss, which is the sum of the photo and collisionally ionized emission components described by Miniutti et al. (2007). A is the cross normalization factor between NuSTAR and XMM ($A = 1.20 \pm 0.05$) and C is the offset between reflected/line components and intrinsic continuum components

---

[15] https://cxc.cfa.harvard.edu/ciao/ahelp/wavdetect.html

[16] https://cxc.cfa.harvard.edu/cal/ASPECT/celmon/

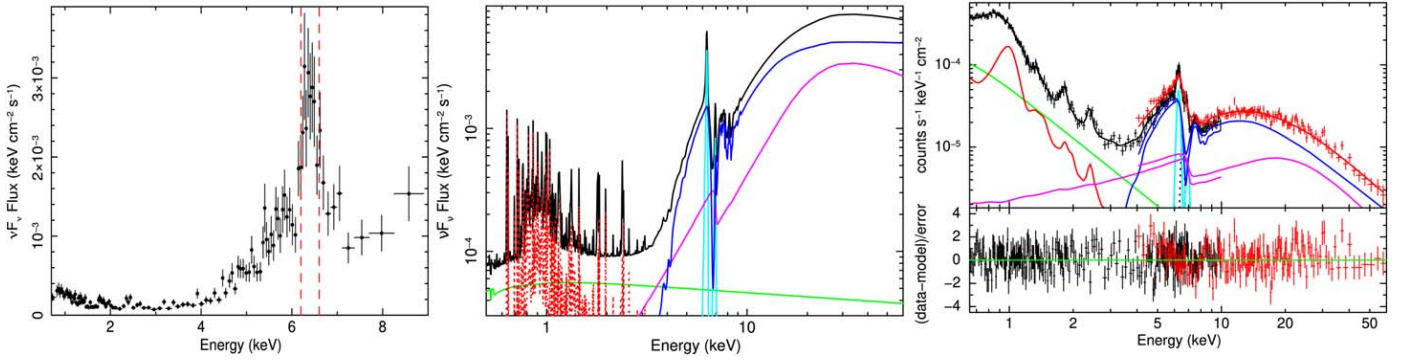

**Figure 6.** Left: Chandra/ACIS spectrum from the inner 2″.5 region of MCG-03-34-64, with the y-axis giving the fluxes in $\nu F_\nu$ units, and the energy of the Fe K$\alpha$ line $E = 6.39 \pm 0.01$. Note the heavy absorption of the soft photons at the lower energies, which is attributed to heavy obscuration of the source. The red dashed lines mark the 6.2–6.6 keV Fe K$\alpha$ band. Center: joint XMM-Newton & NuSTAR spectrum, and MYTorus individual model components. The black curve is the total emission, blue is the transmitted continuum absorbed by Compton-thin matter along the line of sight, magenta is the scattered (reflected) MYTorus component off Compton thick matter, cyan is the Fe K line emission, red is the ionized soft X-ray emission, and green is an unabsorbed power-law component. Right: MYTorus model fit to the joint XMM-Newton and NuSTAR observation. The model components have the same colors as in the center panel. The data and model components are folded through the instrumental responses.

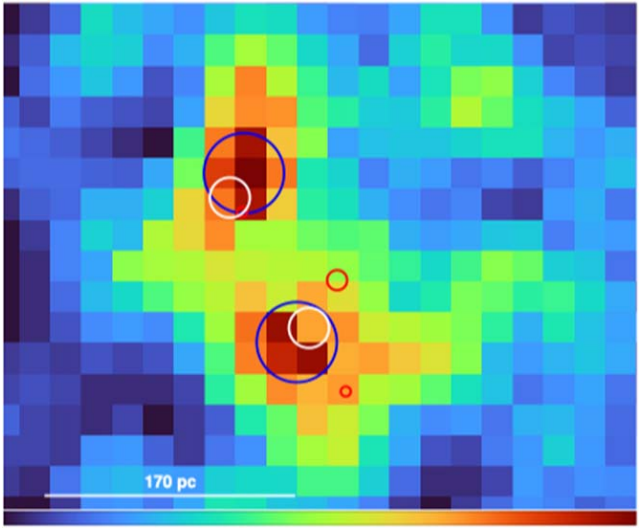

**Figure 7.** Chandra/ACIS-S merged Fe K (6.2–6.6 keV) image of MCG-03-34-64 (one-eighth subpixel, and smoothed with 1 kernel Gaussian). Optical centroids from HST/ACS are shown in red, VLA-A 8.46 GHz centroids in white, and Chandra/ACIS Fe K centroids in blue. The circle sizes reflect uncertainties in the position of the centroids.

($C$ is frozen at 1). The results of the fitting are shown in Figure 6 (center and right panels).

This model returns a good fit, $\chi^2_\nu = 1.05$, with $\Gamma = 2.12 \pm 0.03$, $N_{H_{eq}} = 5.4 \pm 0.3 \times 10^{23}$ cm$^{-2}$ (absorption column density in the equatorial plane, i.e., line-of-sight column density), and $N_H > 6.0 \times 10^{24}$ cm$^{-2}$ (out of line-of-sight column density). In this case, we assume that the hard intrinsic and soft scattered power-law continua have the same photon index.

From the fitting results, we derive an integrated Fe K$\alpha$ luminosity of $L_{(6.2-6.6\text{ keV})} = 1.0 \times 10^{41}$ erg s$^{-1}$ (from MYTL), a scattered (reflected) continuum luminosity of $L_{(2-10\text{ keV})_{\text{scat.}}} = 2.1 \times 10^{41}$ erg s$^{-1}$ (from MYTS), and a scattering corrected intrinsic 2–10 keV luminosity of $L_{(2-10\text{ keV})_{\text{int.}}} = 1.5 \times 10^{43}$ erg s$^{-1}$ (from MYTZ).

Given the quality and resolution of the X-ray observations used in this work, the spectra and fitting models employed in our analysis account for emission within the entire inner region of MCG-03-34-64, and cannot be performed separately for the individual Fe K$\alpha$ peaks uncovered in the Chandra imaging data. In this case, the resulting configuration suggested by MYTorus requires one where one is looking along the edge (Compton-thin line of sight) of a very Compton-thick absorber overall, which obscures both Fe K$\alpha$ regions.

We also note that the difference observed between the summed Fe K$\alpha$ luminosities derived from the Chandra imaging, $\sim 6.4 \times 10^{40}$ erg s$^{-1}$, and the total Fe K$\alpha$ luminosity yielded by the spectral fitting with MYTorus, $L_{(6.2-6.6\text{ keV})} = 1.0 \times 10^{41}$ erg s$^{-1}$, may be attributed to line absorption by the absorber, which in turn implies a higher intrinsic X-ray luminosity, as yielded by the results of the spectral fit.

### 4. Discussion

The results presented in Section 3 reveal puzzling properties of the emission in MCG-03-34-64. Our imaging analysis identified three [O III]-emitting regions in the HST/ACS data, separated by $76 \pm 8$ and $79 \pm 8$ pc (Table 2, Figure 2). In X-rays with Chandra/ACIS, two spatially resolved peaks of emission are observed in the narrow 6.2–6.6 keV Fe K$\alpha$ band, separated by $125 \pm 21$ pc (Table 3, Figure 3). In the radio with VLA-A, Schmitt et al. (2001) previously identified two distinct radio cores in the 8.46 GHz continuum, separated by $116 \pm 14$ pc (Table 4, Figure 5).

Figure 7 shows the Chandra/ACIS Fe K$\alpha$ image and the position of these multiwavelength centroids. The image is binned at one-eighth of the native ACIS-S pixel and smoothed with 1 kernel Gaussian.

#### 4.1. Bolometric Luminosity

Using the [O III] fluxes in Table 2, and $D \sim 78$ Mpc (e.g., de Grijp et al. 1992), we calculate individual [O III] luminosities for each emitting region and for the Sérsic components modeling the HST data, as listed in Table 6. We then calculate individual bolometric luminosities by applying a correction factor of $c = 454$ (Lamastra et al. 2009) to $L_{\text{[O III]}}$ (Table 6). The northern [O III] region has the highest [O III] and bolometric luminosities, $3.5 \times 10^{41}$ erg s$^{-1}$ and $1.6 \times 10^{44}$ erg s$^{-1}$, respectively, while the central region is the least luminous, $1.6 \times 10^{41}$ erg s$^{-1}$ and $7.3 \times 10^{43}$ erg s$^{-1}$. We calculate an integrated bolometric luminosity of $L_{\text{bol[O III]}} = 4.8 \times 10^{44}$ erg s$^{-1}$ by coadding individual values (Table 6).

**Table 5**
Joint XMM and NuSTAR Spectral Fitting Results from MYTorus

| $L_{(6.2-6.6\,\text{keV})}$[a] (erg s$^{-1}$) | $L_{(2-10\,\text{keV})\text{scat.}}$[b] (erg s$^{-1}$) | $L_{(2-10\,\text{keV})\text{int.}}$[c] (erg s$^{-1}$) | $L_{\text{bol}(2-10\,\text{keV})}$[d] (erg s$^{-1}$) |
|---|---|---|---|
| $1.0 \times 10^{41}$ | $2.1 \times 10^{41}$ | $1.5 \times 10^{43}$ | $4.5 \times 10^{44}$ |

**Notes.** We use a correction factor $k = 30$ (Vasudevan & Fabian 2007) to obtain the integrated bolometric luminosity in X-rays.
[a] The integrated Fe K$\alpha$ luminosity as modeled by MYTS.
[b] The scattered (reflected) continuum as modeled by MYTS.
[c] The scattering-corrected intrinsic 2–10 keV luminosity as modeled by MYTZ.
[d] The bolometric luminosity estimated from $L_{(2-10\,\text{keV})\text{int.}}$.

In X-rays, we use the scattering-corrected intrinsic 2–10 keV luminosity from Table 5 and an X-ray correction factor of $k = 30$ (Vasudevan & Fabian 2007) to calculate an integrated bolometric luminosity of $L_{\text{bol}(2-10\,\text{keV})} = 4.5 \times 10^{44}$ erg s$^{-1}$, consistent with $L_{\text{bol[O III]}}$ derived from [O III] (Table 5).

*4.1.1. HST F814W Continuum Fluxes*

The three distinct regions of [O III] emission observed in the continuum-subtracted image (Figures 1 and 2) are also visible in the F814W wide-band image. We estimate the total (integrated) intrinsic flux in the F814W band, $F_{F814W_{\text{total}}}$, assuming a typical spectral energy distribution (SED) as in, e.g., Revalski et al. (2018), Trindade Falcão et al. (2021), and the scattering-corrected intrinsic 2–10 keV luminosity from the MYTorus fit, $L_{(2-10\,\text{keV})\text{int.}} = 1.5 \times 10^{43}$ erg s$^{-1}$. Our results yield a total flux in the band of $F_{F814W_{\text{total}}} \sim 1.0 \times 10^{-11}$ erg s$^{-1}$ cm$^{-2}$.

We also calculate the individual observed fluxes in the F814W band for each [O III] region ($f_{F814W_{\text{obs.}}}$) in the HST/ACS image, assuming the regions fit in the [O III] image (see Table 2) are the same in the F814W band. The F814W image was flux calibrated by multiplying it by $7.104 \times 10^{-20}$ erg cm$^{-2}$ Å$^{-1}$ e$^{-1}$ and the F814W filter bandwidth ($\sim$2710 Å). For simplicity, we repeat the fit presented in Table 2 using Gaussians for the three regions of [O III] emission, which resulted in components with similar parameters and flux differences of less than $\sim$5%. The F814W image was used to measure the peak of emission $f_{F814W_{\text{peak}}}$ at the central position of each corresponding [O III] component, after subtracting the contribution from the host galaxy, which is estimated based on a linear fit to the outer portion of the one-dimensional profile cut through each of the peaks of emission. These $f_{F814W_{\text{peak}}}$ values were converted into $f_{F814W_{\text{obs.}}}$ using

$$f_{F814W_{\text{obs.}}} = 1.1377 \times \text{FWHM}^2 \times (b/a) \times f_{F814W_{\text{peak}}}. \quad (1)$$

Our results reveal that the integrated observed fluxes are $\leqslant$3% of the integrated intrinsic flux in the band (Table 6), with the largest fraction originating from the central region. These fractions are consistent with scattered, hidden continuum (e.g., Pier et al. 1994), but could also include contributions from emission lines (e.g., Kraemer & Crenshaw 2000) and recombination continuum (Osterbrock & Robertis 1985). Therefore, it is unlikely that we will detect AGN continuum emission directly in the optical, and we cannot determine which of these regions harbors the AGN, given that all three regions are consistent with scatter continua from an active nucleus (but see below).

**Table 6**
[O III] Luminosities Calculated from the Measured [O III] Fluxes from Table 2, and Considering a Distance of $D = 78$ Mpc

| Component | $L_{\text{[O III]}}$ (erg s$^{-1}$) | $L_{\text{bol[O III]}}$ (erg s$^{-1}$) | $f_{F814W_{\text{obs.}}}$ (erg s$^{-1}$ cm$^{-2}$) |
|---|---|---|---|
| Northern [O III] | $3.5 \times 10^{41}$ | $1.6 \times 10^{44}$ | $8.3 \times 10^{-14}$ |
| Central [O III] | $1.6 \times 10^{41}$ | $7.3 \times 10^{43}$ | $1.2 \times 10^{-13}$ |
| Southern [O III] | $2.3 \times 10^{41}$ | $1.0 \times 10^{44}$ | $6.4 \times 10^{-14}$ |
| Sérsic | $3.3 \times 10^{41}$ | $1.5 \times 10^{44}$ | … |
| Total | $1.1 \times 10^{42}$ | $4.8 \times 10^{44}$ | … |

**Note** We use a correction factor $c = 454$ (Lamastra et al. 2009) to calculate the bolometric luminosity in each [O III] region. We also show the measured F814W fluxes for each emitting region.

*4.2. Multiwavelength Emission Centroids*

The high fluxes and luminosities found in Section 3 for individual emission regions in the optical, X-ray, and radio bands support the presence of an AGN in this system. However, pinpointing the AGN's location is more challenging. We discuss possible interpretations for the system's configuration, based on our results and the limitations of the data.

*4.2.1. Single AGN+Shocked Interstellar Medium*

One interpretation is that the active nucleus is located at the position of the northern centroids ([O III], Fe K$\alpha$, and radio; see Figure 7), based on the fluxes of individual components (Tables 2, 3, and 4). In this single AGN scenario, the remaining emission centroids (central Fe K$\alpha$, [O III] and radio centroids, and southern [O III]) may arise from the interaction between the AGN and the interstellar medium (ISM). This would manifest as a mix of photoionized and collisionally ionized (shocked) gas from an extended NLR, similar to NGC 3393 (e.g., Maksym et al. 2016). Such an interpretation is consistent with previous spectral fitting results for this galaxy, which indicate a mix of photoionized and shock-ionized gas in the soft X-ray emission (Miniutti et al. 2007). Similarly, it is possible that the AGN in this system is located at the position of the central [O III], Fe K$\alpha$, and radio peaks, while the remaining multiwavelength centroids may be attributed to AGN–ISM shock emission.

However, the "single AGN+shocked ISM" scenario in this galaxy is challenged by the high and nearly equal X-ray luminosities derived for each Fe K$\alpha$ region in the Chandra data. The 6.4 keV Fe K$\alpha$ emission line typically arises from irradiation of the neutral (or low ionized) iron by a hard X-ray source, and in AGNs, this line may originate from the innermost part of the accretion disk, close to the central SMBH. This emission process is known to be extremely energetically inefficient, with an overall conversion factor of $\sim$1% of the intrinsic X-ray luminosity (e.g., Murphy & Yaqoob 2009). This is illustrated by the fitting results with MYTorus (Section 3.2, Table 5), in which the total Fe K$\alpha$ luminosity $L_{(6.2-6.6\,\text{keV})}$ is $\sim$0.7% of the total scattering-corrected intrinsic 2–10 keV luminosity $L_{(2-10\,\text{keV})\text{int.}}$.

In Section 2.2, the derived Fe K$\alpha$ luminosities for the individual regions are on the order of $3.2 \pm 0.6 \times 10^{40}$ erg s$^{-1}$ (Table 3). This suggests that two individual hard continuum sources are needed to provide the required energy input (George & Fabian 1991; Murphy & Yaqoob 2009). For reference, an extended Fe K$\alpha$ region in NGC 5728 ($D\sim41$ Mpc) was reported with an integrated luminosity of $L_{(6.1-6.6\,\mathrm{keV})} \sim 6.7 \times 10^{39}$ erg s$^{-1}$ within a $1''\!.5$–$8''$ annular region (excluding the nucleus; Trindade Falcão 2023, 2024). This luminosity is 1 order of magnitude lower than those of the individual Fe K$\alpha$ peaks in MCG-03-34-64 (Table 3). In contrast, a luminosity of $L_{(6.1-6.6\,\mathrm{keV})} \sim 6.0 \times 10^{40}$ erg s$^{-1}$ was found within the inner $1''\!.5$ of NGC 5728 (Trindade Falcão 2023, 2024), comparable to the luminosities of the individual Fe K$\alpha$ peaks in MCG-03-34-64. While the emission in NGC 5728 was attributed to shocked gas in the ISM (Trindade Falcão 2023, 2024), a similar origin in MCG-03-34-64 seems unlikely due to the high luminosities of the individual regions. Therefore, it is unlikely that collisional ionization in the NLR alone powers the individual Fe K$\alpha$ peaks observed in MCG-03-34-64.

Table 7
Distances between Multiwavelength Centroids Found in This Work: HST/ACS, VLA-A, and Chandra/ACIS-S (Fe K$\alpha$)

| Components | Distance (arcsec) | Distance (pc) | Distance Range (pc) |
|---|---|---|---|
| Northern [O III] → central [O III] | $0.233 \pm 0.02$ | $79 \pm 8$ | 71–87 |
| Central [O III] → southern [O III] | $0.223 \pm 0.02$ | $76 \pm 8$ | 68–84 |
| Northern [O III] → southern [O III] | $0.413 \pm 0.01$ | $140 \pm 5$ | 135–145 |
| Northern radio → central Radio | $0.338 \pm 0.04$ | $116 \pm 14$ | 102–130 |
| Northern Fe K$\alpha$ → central Fe K$\alpha$ | $0.369 \pm 0.08$ | $125 \pm 21$ | 104–146 |
| Central [O III] → central Fe K$\alpha$ | $0.132 \pm 0.06$ | $45 \pm 21$ | 24–66 |
| Southern [O III] → central Fe K$\alpha$ | $0.107 \pm 0.06$ | $36 \pm 21$ | 15–57 |

The energetic mismatch is strengthened by the similar scattering corrected intrinsic X-ray luminosities calculated for these galaxies, $L_{(2-10\,\mathrm{keV})\mathrm{int.}} = 1.5 \times 10^{43}$ erg s$^{-1}$ for MCG-03-34-64 (Table 5), versus $L_{(2-10\,\mathrm{keV})} \sim 1.2 \times 10^{43}$ erg s$^{-1}$ for NGC 5728 (Zhao et al. 2021), suggesting that the differences in Fe K$\alpha$ luminosities between the two objects cannot be attributed solely to differences in the intrinsic luminosity of the nuclear source.

#### 4.2.2. Dual AGN+Shocked ISM

Following the discussion in Section 4.2.1, the high Fe K$\alpha$ luminosities (Table 3), and the high energies required for the production of such line emission suggest that both Fe K$\alpha$ regions could be powered by an active SMBH. In this scenario, the northern emission centroids would pinpoint the location of one AGN, while the central emission centroids (radio, Fe K$\alpha$, and optical) may be associated with a second active SMBH in this system (given the high fluxes found for the central optical region in the F814W continuum band; Table 6). The distances measured between the different centroids attributed to each AGN are consistent across different wave bands (Table 7), supporting the dual AGN scenario.

In this scenario, the southern [O III] region may arise from collisionally ionized emission in the ISM. Shock emission from jet–ISM interaction at the southern optical centroid location is supported by (1) the morphology of the radio emission at the central radio peak, which is observed to bend southward (Schmitt et al. 2001), in the direction of the southern [O III] peak (see Figures 5 and 3, and Section 3.1.3); and (2) the morphology of the Chandra 0.3–3 keV (soft) emission, which is also observed to bend southward in the direction of the southern [O III] centroid (see Figure 3, and Section 3.1.3). The lack of a corresponding southern radio or hard X-ray counterpart is consistent with the hypothesis that the northern and central [O III] centroids are powered by individual AGNs.

The dual AGN scenario is strengthened by the consistent values between the estimated bolometric luminosities derived from [O III] and X-rays (Tables 5 and 6), and the detection of nearly equal Fe K$\alpha$ emission peaks in the Chandra image, a powerful tool for identifying and confirming dual AGN systems (De Rosa et al. 2022). In the 3–5 keV Chandra image (Figure 3), the northern nucleus appears brighter than the central nucleus, although some extended emission is observed toward the central nucleus in this band. The column densities in the transmission of the two nuclei may not be the same, i.e., the AGN located at the central Fe K$\alpha$ region could have a higher absorbing column and appear fainter at lower energies. A higher contribution from the photoionized+thermal extended soft X-ray gas is expected at these lower energies. Given the resolution of the analyzed X-ray data, performing a separate spectral analysis of each individual Fe K$\alpha$ region is currently impractical.

### 5. Summary and Conclusions

We analyze new HST/ACS and Chandra/ACIS observations of the nearby Seyfert galaxy MCG-03-34-64, along with archival HST/ACS, XMM-Newton/Epic-pn, NuSTAR, and VLA-A data sets. Our analysis reveals the following:

In the optical with HST: Three emission regions are uncovered in the narrow-band continuum-subtracted [O III] and F814W continuum images (Figures 1, 2), separated by $79 \pm 8$ pc (northern to central), and $76 \pm 8$ (central to southern). Diffraction spikes from the bright [O III]-emitting regions suggest individual sizes of $\lesssim 60$ pc. We derive individual bolometric luminosities in the range $L_{\mathrm{bol[OIII]}} \sim 0.73 - 1.6 \times 10^{44}$ erg s$^{-1}$, and an integrated bolometric luminosity $L_{\mathrm{bol[O\,III]}} = 4.8 \times 10^{44}$ erg s$^{-1}$ from [O III] (Table 6).

In X-rays with Chandra: Two spatially resolved emission centroids are detected in the 6.2–6.6 keV Fe K$\alpha$ image, separated by $125 \pm 21$ pc. These peaks are evident in individual exposures and the merged data set. The northern and central Fe K$\alpha$ regions have $36 \pm 6$ and $37 \pm 6$ counts in the narrow 6.2–6.6 keV band, respectively, corresponding to $\geqslant 6\sigma$ detections, and nearly equal Fe K$\alpha$ luminosities, $L_{(6.2-6.6\,\mathrm{keV})} \sim 3.2 \pm 0.6 \times 10^{40}$ erg s$^{-1}$.

In X-rays with XMM and NuSTAR: Fitting the joint spectrum with a physically motivated model yields an integrated Fe K$\alpha$ luminosity of $L_{(6.2-6.6\,\mathrm{keV})} = 1.0 \times 10^{41}$ erg s$^{-1}$, a scattered (reflected) continuum luminosity of $L_{(2-10\,\mathrm{keV})\mathrm{scat.}} = 2.1 \times 10^{41}$ erg s$^{-1}$, and a total, scattering corrected intrinsic 2–10 keV luminosity of $L_{(2-10\,\mathrm{keV})\mathrm{int.}} = 1.5 \times 10^{43}$ erg s$^{-1}$.

In the radio with VLA: Two emission regions are observed in the 3.6 cm VLA continuum image (Schmitt et al. 2001), spatially colocated with the northern and central Fe K$\alpha$ and [O III] regions.

We propose two possible physical interpretations of our results, and discuss these in the context of our analysis:

1. The "single AGN+shocked ISM" scenario, which proposes the existence of a single active nucleus in the system, while the remaining multiwavelength centroids may be attributed to the interaction of the ISM with the radio jet in the NLR.

This scenario is strengthened by:

a. Previous X-ray studies on this source, which find evidence for a mix of collisionally and photoionized X-ray gas in the NLR (Miniutti et al. 2007).

This scenario is challenged by:

b. The high Fe K$\alpha$ luminosities derived for individual regions and the energies required for the production of such line emission.

2. The "dual AGN+shocked ISM" scenario, which proposes the existence of a dual SMBH pair in this system separated by just $125 \pm 21$ pc.

This scenario is strengthened by:

a. The detection of two spatially resolved Fe K$\alpha$ regions in the Chandra imaging data, with high individual luminosities (Table 3).

b. The detection of three very bright and compact (<60 pc diameter) [O III]-emitting regions in the HST imaging data, and the respective individual bolometric luminosities (Table 6).

c. The detection of spatially coincident Fe K$\alpha$, radio, and [O III] centroids at the northern and central regions. This is the first time spatially resolved, multiwavelength emission centroids in X-rays, radio, and optical are detected colocated in a nearby candidate dual AGN. For comparison, the recent study of Koss et al. (2023), which identified the presence of a dual AGN system separated by ∼230 pc in UGC 4211, detected colocated optical (HST F814W, MUSE AO [O III], and H$\alpha$), NIR (Keck J and K'), and submillimeter (Atacama Large Millimeter/submillimeter Array continuum at ∼230 GHz) centroids at the position of the two nuclei, but with no confirmation from X-rays or radio observations.

In summary, although we cannot definitively confirm or exclude the physical scenarios presented here, identification of the two nuclei in a deeper Chandra exposure would help to confirm a possible dual black hole system in this galaxy. Analysis of gas kinematics in the nuclear region of MCG-03-34-64 is crucial to determine the nature of the observed structures. Kinematic information obtained with HST/STIS long-slit spectroscopy could reveal disturbed kinematics expected from either the individual outflows of two SMBHs or the highly disturbed kinematics resulting from the merger environment. This information cannot be obtained from the archival X-Shooter data and requires the resolution of HST to probe the ∼100 pc region of interest.

## ORCID iDs

Anna Trindade Falcão https://orcid.org/0000-0001-8112-3464
T. J. Turner https://orcid.org/0000-0003-2971-1722
S. B. Kraemer https://orcid.org/0000-0003-4073-8977
J. Reeves https://orcid.org/0000-0003-3221-6765
V. Braito https://orcid.org/0000-0002-2629-4989
H. R. Schmitt https://orcid.org/0000-0003-2450-3246
L. Feuillet https://orcid.org/0000-0002-5718-2402